\documentstyle[12pt,aasms4,psfig]{article}

\newcommand{\solm}{M$_{\odot}$}

\newcommand{\solar}{L$_{\odot}$\ }

\newcommand{\rf}{\par\noindent\hangindent 15pt {}}
\newcommand{\apjj}[2]{Ap. J., #1, #2.}
\newcommand{\apjjs}[2]{Ap. J. Supp., #1, #2.}
\newcommand{\apjjl}[2]{Ap. J. (Letters), #1, #2.}
\newcommand{\asa}[2]{Astron. Astrophys., #1, #2.}
\newcommand{\arasa}[2]{Ann.Rev.Astr.Ap., #1, #2.}
\newcommand{\asas}[2]{Astron. Astrophys. Suppl., #1, #2.}
\newcommand{\anrev}[2]{Ann.Rev.Astron.Astrophys., #1, #2.}

\newcommand{\mn}[2]{M.N.R.A.S., #1, #2.}
\newcommand{\ajj}[2]{A. J., #1, #2.}

\newcommand{\vol}[1]{1}
\newcommand{\nhico}{$\frac{N_{H_2}}{I_{CO}}$}

\lefthead{Schinnerer et al.}
\righthead{The Host Galaxy of the QSO I~Zw~1}

\begin{document}

\title{
 Molecular Gas and Star Formation in the 
\\
Host Galaxy of the QSO I~Zw~1}

\author{
E. Schinnerer, A. Eckart, L.J. Tacconi }

\affil{ Max Planck Institut f\"ur extraterrestrische Physik, 85740 Garching, Germany}
\vspace*{10cm}
to appear in Astrophysical Journal (accepted)

\begin{abstract}
We have investigated the ISM of the I~Zw~1 QSO host galaxy with Plateau de Bure 
mm-interferometry and high angular resolution near-infrared imaging spectroscopy. We  
have detected a circumnuclear 
gas ring of diameter $\sim$ 1.5'' ( 1.8 kpc) in its millimetric CO line emission and have
mapped the disk and the spiral arms of the host galaxy 
in the $^{12}$CO(1-0) line at 115 GHz as well as in the H (1.65 $\mu$m)
and K (2.2 $\mu$m) band .
Combining our new mm- and NIR-data with available estimates of the radio- and
far-infrared contributions to the nuclear emission, we find strong evidence for a 
nuclear starburst ring. A comparison to other sources with nuclear activity indicates 
that these rings may be a common phenomenon and contribute a large fraction of the 
central luminosity.
\\
Both the CO rotation curve as well as the NIR and optical images are consistent with
an inclination of (38$\pm$5)$^o$. Using this we obtain a total
dynamical mass of (3.9$\pm$1.6)$\times$10$^{10}$ \solm ~and a cold molecular gas mass of 
(7.5$\pm$1.5)$\times$10$^9$ \solm ~for the inner 3.9~kpc. With an estimate
of the nuclear stellar contribution to the mass and light from NIR spectroscopy and
assuming that the contribution of the HI gas to the overall mass of the inner 3.9~kpc
is small we derive an \nhico-conversion
factor close to 2$\times$10$^{20}$$cm^{-2}K^{-1}km^{-1}s$ found
for molecular gas in our Galaxy and many nearby external galaxies.
\\
A comparison to broad band spectra of spiral galaxies, ellipticals 
and the nucleus and disk in NGC~7469 suggests bluer disk colors for
I~Zw~1, and
that star formation in the host galaxy and the western companion of
I~Zw~1 is enhanced. This is also supported by a starburst analysis using all
available data on the northwestern spiral arm.
The presence of molecular material within the disk and on the 
arm indicates that at least in this region 12~kpc from the nucleus
star formation, and not scattered light from the QSO nucleus, is 
responsible for the blue disk colors.
\end{abstract}


\section{INTRODUCTION}

Recent investigations of bright active galactic nuclei 
have shown that the starburst phenomenon contributes a major 
fraction of the total luminosity of these sources (Genzel et al. 1997).
In addition several
models suggest evolutionary links between 
different classes of extragalactic sources
(e.g. Osterbrock 1993, Norman and Scoville 1988, Sanders et al. 1988, 
Rieke, Lebofsky \& Walker 1988). 
A key test of these evolutionary sequences is to investigate the 
structure and concentration of the molecular material 
as well as the distribution and composition of the stellar population
in the nuclei and circumnuclear regions of a sample of galaxies representing the different classes. 
Several luminous and ultra-luminous extragalactic nuclei have
been observed with high angular resolution, and have shown the presence of circumnuclear
starburst rings. 
These rings, therefore, deserve special attention since they are
important tracers of the dynamics and star formation in active and
luminous galactic nuclei.
Due to the combined requirement of high spatial resolution and sensitivity
these investigations become increasingly difficult when going to larger
distances.
In the course of such a systematic effort we present new 
results we have obtained with imaging and spectroscopy on the nearby QSO I~Zw~1.
\\
\\
I~Zw~1, at a redshift of z=0.0611 (Condon et al. 1985), is thought to be the 
closest QSO because of its high optical nuclear
luminosity (M$_V$=-23.8$^{mag}$ V\'{e}ron-Cetty \& V\'{e}ron 1991). 
For this redshift its distance is 
244 Mpc
assuming H$_o$=75 kms$^{-1}$Mpc$^{-1}$. I~Zw~1 shows 
spectral properties of high-redshift QSOs (e.g. CIV ($\lambda$
1549$\AA$) which are blueshifted by 1350 km/s (Buson \& Ulrich 1990)).
It also has a high X-ray luminosity 
(Kruper et al. 1990, 
Boller, Brandt \& Fink 1996).
I~Zw~1 is the prototype object of the Narrow-Line Seyfert 1 (NLS1)
class. The FWHM of the H$\beta$ line is $\approx$ 1240 km/s, the
OIII/H$\beta$ ratio is smaller then 3 and strong optical FeII
multiplets are detected (Halpern \& Oke 1987).
The host galaxy disk has been detected in the 
V, R and H band (Bothun et al. 1984, Hutchings \& Crampton 1990,
McLeod \& Rieke 1995), and its molecular gas has been observed in the
$^{12}$CO(1-0) (Barvainis et al. 1989), $^{12}$CO(2-1) and
$^{13}$CO(1-0) lines (Eckart et al. 1994). 
The two objects seen near the edges of the disk are a foreground star  
to the north, and a companion galaxy to the west (Hutchings \& Crampton 1990).
The QSO nucleus is located in a gas rich host galaxy disk, which  
makes I~Zw~1 an ideal candidate for studying the properties of QSO hosts.
\\
\\
We present the observations and data reduction in
section 2 and discuss the distribution, mass and dynamics of the
molecular gas in section 3.
The near-infrared imaging and spectroscopy results are given in 
section 4,
and the possibility of a massive nuclear starburst (section 5) is then
followed by a comparison with other starburst rings (section 6).
In section 7 we compare the optical/near-infrared broad band spectra
of the northwestern spiral arm in the I~Zw~1 host galaxy
and the western companion to those of spiral galaxies, ellipticals and
the disk and nucleus of NGC~7469.
We give a summary and conclusions in section 8.

\section{OBSERVATIONS AND DATA REDUCTION}

\subsection{Millimeter-Spectroscopy}

We have mapped the $^{12}$CO J=1-0 line emission in I~Zw~1 in January/February
1995 with the IRAM
millimeter interferometer on the Plateau de Bure, France 
(Guilloteau et al. 1992).
The four 15~m antennas were positioned in four different configurations,
providing 24 baselines, ranging from 24~m to 288~m length. The antennas were equipped
with SIS receivers with single-sideband (SSB) system temperatures
above the atmosphere of 170 K. The observed frequency was 108.633 GHz
due to the redshift z=0.0611 of I~Zw~1 (Condon et al. 1985). 3 C 454.3 was observed for bandpass calibration,
while phases and amplitudes were calibrated on 0106+013.
\\
The CO maps were made using the IRAM reduction package CLIC, and were CLEANed
with the IRAM reduction package GRAPHIC.
The resolution of the synthesized beam was 1.9'' (uniform weighting),
but we made 5'' resolution CLEAN maps (natural weighting) with spectral 
resolutions of 10 km/s and 40 km/s
to study the extended disk structure and velocity field. 
For the studies of the core 
component we used the 1.9'' resolution CLEANed maps with a spectral resolution of 20 km/s.
To investigate the dynamics of the nucleus we calculated velocity maps as well as
p-v diagrams along the major and minor kinematic axis of I~Zw~1.

\subsection{Near-Infrared Spectroscopy and Imaging}

I~Zw~1 was observed in the K band (2.20 $\mu$m) in January 1995
with the MPE imaging spectrometer 3D (Weitzel et al. 1996, Thatte et al. 1995)
at the 3.6m telescope in Calar Alto, Spain.
The observations in the H band (1.65 $\mu$m) were carried out in December 1995 
at the William Herschel Telescope in La Palma, The Canary Islands. 
\\
For both sets of observations the image scale was 0.5''/pixel.
The total integration time on source was 4200~s for the
K band and 1530~s for the H band.
\\
3D obtains simultaneous spectra (R$\equiv\lambda/\triangle\lambda\approx$1000
for H band and 750 for K band) of an 8'' $\times$ 8'' field.
This is done using an image slicer which rearranges
the two-dimensional focal plane onto a long slit of a grism. The dispersed
spectra are then detected on a NICMOS~3 array.
A detailed description of the instrument and the data reduction are given in Weitzel
et al. (1996).
The data reduction procedure converts each two dimensional image into a
three dimensional data cube with two spatial axes and one spectral axis.
The data cubes are coadded after rebinning onto a 0.25'' grid and centered on the
continuum peak. All images are dark-current and sky-background subtracted,
corrected for dead and hot pixels, and spatially and spectrally flatfielded.
To correct the effects of the Earth's atmosphere to the K band spectrum, a
standard star was observed. This standard spectrum was first 
divided by a template spectrum of the same spectral type 
(Kleinmann \& Hall 1986) in order to remove stellar features.
For the region of the central (1.522 - 1.714~$\mu$m) H band we used a model atmosphere, 
because currently no H band template spectra are available at our spectral resolution.
The model atmosphere for the atmospheric correction in the H band was calculated via ATRAN (Lord 1992).
At the H band edges, however, the model atmosphere spectrum was substituted  
with the atmospheric transmission derived via the standard star, since
this is probably more reliable,
and the standard star does not show strong features in these
wavelength ranges.
The effect due to different zenith distance of source and standard star
in H and K band was minimized using the ATRAN atmospheric model (Lord 1992), mainly to
correct for the different atmospheric absorption.
The source data were divided by the atmospheric
transmission spectrum.
We flux calibrated the data adopting the integrated flux density values
given by Edelson \& Malkan (1986) for 5'' circular apertures centered
on the nucleus. 
\\
\\
With a larger field of view the H and K band continuum emission of the QSO host galaxy
was observed in July 1995 with the ESO IRAC2 camera at the ESO/MPG 2.2m telescope
on La Silla, Chile. 
The image scale was 0.27'' per pixel resulting in a 69''$\times$69'' field of view. 
We applied a correction for the sky background, dead pixels and a flat field to the data.
The different frames were coadded using
the central peak of the brightness distribution as a reference.
Again we adopted the calibration used by Edelson \& Malkan (1986)
as described above.
As determined from images of the compact nucleus, the seeing was about 1.3''. For a better
comparison to the interferometer maps and to improve the signal-to-noise we convolved
the images to a resolution of $\sim$2.4''.

\section{THE PROPERTIES OF THE MOLECULAR GAS}

Knowledge of the properties of the molecular gas is essential to the understanding
of star formation in and the fueling of AGNs, since molecular clouds are the major
reservoir for these phenomena. Interferometry at mm-wavelengths
allows high angular resolution dynamical studies of the molecular
gas in the vicinity of the active nucleus. 
For the first time we have detected the molecular line emission in the spiral arms of a QSO host
galaxy. In addition we are able to decompose the line emission
into a core and disk component.
Analyzing the velocity field we have found a circumnuclear ring of
molecular gas with similar size to starburst rings found in nearby galaxies. 
At our spatial resolution of 1.9'' (2.2~kpc) we see no obvious
sign of gas streaming directly into the very nucleus.

\subsection{The Distribution of the Molecular Gas in I~Zw~1}

Since the molecular gas is the reservoir for the star formation, it is important to
investigate its distribution throughout the entire host galaxy. This can shed light on
any differences which exist between the star forming processes in the host galaxy disk and 
those of the nuclear region.
Therefore we made integrated CO emission maps (Fig.1) at spatial resolutions of 1.9'' and 5.0'',
using only emission exceeding the 3$\sigma$ level in each channel. 
The nuclear peak in the CO emission shows an offset of 1'' compared to the VLA 
coordinates, probably due to differences of the interferometer phase centers of the PdBI 
and the VLA. 
\\
In the 5'' spatial resolution map we detect two spiral arms to the west and east 
of the nucleus (Fig.1 and 2 (Plate~1)). 
In the high resolution map these arms are apparent at the lowest contour level.
The extended line emission from the host galaxy is resolved out at high resolution.
The northwestern spiral arm has also been seen in
optical and infrared broad-band images (see section 7).
We also find a smooth disk with an extent of 20'' EW and 13'' NS. 
As shown in Fig.1 most of the flux originates in the central 9''. 
In the 5'' resolution map the nucleus has a FWHM of 6'', which results 
in an intrinsic diameter of about 3.3'' after deconvolution with the CLEAN beam 
at a spatial resolution of 5''. In the 1.9'' resolution map the FWHM diameter of
the core is about 3'', resulting in an intrinsic diameter of 2.3''. 
Also the integral in a 5.5''$\times$5.5'' box centered on the nucleus in the 
5'' resolution map equals 10 Jy~km~s$^{-1}$  whereas the integral in a 2''$\times$2''
box in the 1.9'' resolution map results in a flux of about 5 Jy~km~s$^{-1}$.
The two measurements clearly indicate the presence of a more extended 
component, which is partially resolved in both maps.
This shows that the interferometer data may not contain all of the total
flux of the source.
This is also supported by a comparison to IRAM 30~m data as shown in the following 
section.

\subsubsection{Decomposing the $^{12}$CO Line Emission}

We have compared the
spectrum taken by Barvainis et al. (1989) with the IRAM 30m
telescope with a total spectrum of our IRAM PdB Interferometer observations 
to derive the overall extended line flux emitted by the disk of the I~Zw~1 host galaxy.
Both spectra show a double horned profile, with the 30m spectrum having
more flux in the horns. 
In double horned line profiles the emission in the horns is dominated by an extended
disk component. In addition the 30~m beam size is smaller than the optical disk size and 
of the same order as (or as well smaller than) the structure of the CO line emission 
measured with the PdB interferometer.
This indicates that our observations as well as the 30~m
observations have missed some flux from the disk. The amount of flux missed by the
two observations depends on the disk size, shape and assumed model (see below and
section 3.1.2). To decompose the flux into a nuclear and a disk component we estimated 
the contribution of the compact 3.3'' FWHM core (we measured with the PbBI) to the
total flux (0.078 Jy, value measured between the two horns) of the 30~m telescope 
(Barvainis et al. 1989). From our low 
resolution $^{12}$CO (1-0) map we measure a FWHM of the nuclear component
of about 3.3'' and a peak flux of 0.040 Jy close to the systemic velocity of
I~Zw~1. As defined below we used correction factors F$_G$ and F$_D$ that have to
be applied to the measured flux
for the beam filling under the assumption that beam and source are both
circular and the beam has a Gaussian profile (Dickel 1976). 
\\
\\
If the intensity distribution of the source is Gaussian, we get
\\
\begin{equation}
F_G = \frac{(\theta_{FWHM})^2}{(\theta_B)^2} = \frac{(\theta_B)^2 + (\theta_S)^2}{(\theta_B)^2} = 1 + (\frac{\theta_S}{\theta_B})^2
\end{equation}
\noindent
where $\theta_S$ and $\theta_B$ are the FWHM of the source and the beam.
If the intensity distribution of the source is flat and disk-like (Heeschen 1961), we obtain
\\
\begin{equation}
F_D = \frac{ln2 \cdot (\frac{\theta_S}{\theta_B})^2}{1 - exp(-ln2 \cdot (\frac{\theta_S}{\theta_B})^2)}
\end{equation}
\noindent
where $\theta_S$ is the diameter of the uniform disk and $\theta_B$ is the FWHM of the beam.
Using the beam filling correction factors calculated from the known PdBI and 30m beams
and assuming a Gaussian nuclear source size of 3.3'' we find that the nucleus accounts
for 0.056~Jy, i.e. 70\% of the total line flux measured by the 30m telescope.
We, therefore, decomposed the
30m line flux into a nuclear (0.056 Jy) and a disk (0.022 Jy) component. 
\\
The decomposition is consistent with the source extents and structures found in the
channel maps (Fig.1, 2 and 3), and in optical and NIR-images.
Alternatively the PdBI and 30m spectra only match if we assume unrealistically high
calibration errors $\geq$~20 \% for both observed values, with the 30~m
flux being too high and the PdBI flux being too low.

\subsubsection{Size of the Host Galaxy in the $^{12}$CO Line Emission}

Assuming a disk flux of 0.022~Jy, we derive
its size, since we have two measurements at different angular resolutions.
For the disk we assume
both Gaussian and flat intensity distributions. 
The value of the background noise in the single channel maps is of the order of 
0.006~Jy (3$\sigma$ level). This value was used together with the disk flux to
estimate a lower limit to the spatial extent of the disk via the beam filling
factors.
For a Gaussian (flat) flux distribution we get a disk diameter of 9'' (12'') 
or higher taking into account that parts of the extended emission might also
have been missed by the 30~m measurements.
This is in good agreement with the upper limit in disk size of about
30'' obtained from optical and NIR observations.
This is especially true if we consider the extended wings in case of a
more realistic Gaussian distribution.
The molecular disk size of the host galaxy should have an extent of
about 10~kpc (or larger), compared to that of the bright core component with a 
FWHM of 3.3'' or about 3.9 kpc.

\subsection{The Dynamics of the Molecular Gas in I~Zw~1}

Our $^{12}$CO(1-0) spectrum shows a double-horn profile characteristic of rotating disks,
and is similar to the CO spectra of I~Zw~1 presented in Barvainis et al. (1989) and 
Eckart et al. (1994).
The rotation of the extended disk emission is obvious in the channel maps (Fig.3).
From the mean velocity field (Fig.4) we measure a kinematic major axis position
angle of 135$^o$. The velocity field shows no signs of a very strong
interaction with the western companion. This companion is not detected at our S/N level
of 0.006~Jy.
\\
In Fig.5 we present the position-velocity diagrams along the major kinematic axis
for the two different spatial resolutions. Each half of the p-v diagram
represents the projected rotation curve for I~Zw~1.

\subsubsection{Inclination}

The exact inclination of the I~Zw~1 host galaxy is not known, 
although the optical and NIR-images indicate a fairly low inclination of 
$\sim$~40$^o$ (Bothun et al. 1984).
To test this value for the inclination kinematically we assumed a typical disk rotation
velocity of (250$\pm$25) km/s (Rubin et al. 1985) for the flat part of the rotation 
curve (at a distance of about 10'') assuming this is characteristic of motions in the disk.
From this we derive an inclination in the range of (34$\pm$4)$^o$ 
depending on the chosen rotation velocity. 
This value is in approximate agreement with the shape of the 50\% contour lines in the 
maps in Fig.1 under the assumption that the
CO line emission has an intrinsically azimuthally symmetric distribution.
As a compromise we have chosen a value of 
(38$\pm$5)$^o$  which would be consistent with a rotation velocity of (230$\pm$30) km/s
for the disk. 
This value, as well as the
position angle of the major axis of 135$^o$ that we get from the CO velocity field are 
in agreement with the observed ellipticity of the R band image (Hutchings \& Crampton 1990)
assuming an intrinsically circular disk.

\subsubsection{pv-Diagrams}

The p-v diagrams show evidence for a circumnuclear ring, extended disk emission and the
presence of a large molecular cloud complex located close to the nucleus. In the
following we will summarize observational properties of these regions:

\noindent
{\em The circumnuclear ring:} 
The pv-diagram (Fig.5a) made at 1.9'' angular resolution has two flux peaks at 
velocities of $\pm$180 km/s, indicating the presence of a nuclear ring or spiral arms, 
with a diameter of about 1.6 ''. 
The nuclear velocity dispersion was measured in a pv-diagram 
along the kinematic minor axis. The values given below are the 50\% contour widths 
corrected for the spectral resolution. The intrinsic velocity dispersion is estimated 
assuming a Gaussian velocity distribution. 
The nuclear source exhibits a high velocity dispersion of $\sigma$=350 km/s,
assuming that the gas is isothermal
and that the velocity dispersion is radius independent for radii less than 2''.
This is also clearly demonstrated and supported by the 
pv-diagrams along the minor axis (Fig.5).
However, the $\sigma$ derived  above may not only be due to 
pure cloud-cloud velocity dispersion but may also contain contributions
from a more complex velocity field like a tilted ring system as in 
Centarus~A (Nicholson et al. 1992, Quillen et al. 1992, 1993,
Sparke 1996) or streaming motions.
\\
{\em The GMC complex:} The single emission peak at a
velocity of 220 km/s and approximately 2''E and 2''S of the nucleus could be due to 
a molecular gas complex located in or near the nuclear ring. 
This GMC complex also appears in the low spatial resolution map (Fig.5c).
It seems to be kinematically unresolved, indicating a small velocity dispersion 
of about 20 km/s, similar to what is observed for GMC complexes in our Galaxy. 
The molecular mass contained in this complex 
(estimated from the CO luminosity read off from the pv-diagrams in Fig.5)
is $\sim$ 5 \% of the total nuclear molecular gas mass. This is one order of 
magnitude higher than that measured for GMC complexes in our Galaxy.
\\
{\em The disk emission:}
The flat part of the rotation curve levels out at velocities of $\pm$140 km/s. Our S/N
ratio enables us to trace the extended disk emission out to radii of 18~kpc.
The velocity dispersion of the disk component is $\sim$ 17 km/s.
\\
{\em Modelling the p-v diagrams:}
We have calculated model pv-diagrams to improve our understanding of the 
structure and dynamics of the nucleus and the disk of I~Zw~1 (see Fig. 5). 
The model projects a given galaxy flux geometry onto a sky grid taking 
inclination and tilt into account. To this flux geometry a rotation curve is
applied. The beam size and velocity resolution of the observation are then used
to calculate the model pv-diagram. To model the molecular gas emission in I~Zw~1 we assumed 
a two component Gaussian flux distribution. One Gaussian had
a FWHM of 10'' centered at a radius of 4'' and the other a
FWHM of 1.65'' centered at a radius of 0.8''. The ratio between the two Gaussian
flux peaks was 50 with the inner one being the brighter one. This model flux distribution 
is plotted in Fig.6. 
\\
The azimuthally symmetric modelled p-v diagrams match the measured ones very well, 
except that they do not reproduce emission from the single GMC complex. Also they
account only approximately for the emission of the disk component. The remaining
differences between the data and the model are due to the presence of spiral
arms in the observed maps (see Fig.1). The modelled
p-v diagrams do demonstrate the need of a circumnuclear ring-like distribution of the
molecular gas, as well as the need for a high velocity dispersion in the nuclear source.
This high velocity dispersion could be due to gas streaming along a bar, because the
nuclear emission in the velocity field (Fig.4) is slightly S-shaped as one would expect 
for streaming (Sanders \& Tubbs 1980, Tacconi et al. 1994).

\subsection{The Mass of the Molecular Gas in I~Zw~1}

Since the molecular gas mass is a substantial part of the dynamical mass,
it is essential to know its distribution and mass.
With velocity widths of 380 km/s for the nucleus and 310 km/s 
(distance horn-to-horn)
for the disk and a $\frac{N_{H_2}}{I_{CO}}$-conversion factor
of 2.0$\times$10$^{20}$$\frac{cm^{-2}}{K kms^{-1}}$ (Strong et al. 1987), 
we get a molecular gas mass for the nuclear component of 
7.5$\times$10$^9$M$_{\odot}$, and of 2.7$\times$10$^9$M$_{\odot}$ 
for the Gaussian disk and 0.7$\times$10$^9$M$_{\odot}$ for the flat
disk (see Table 1 and 2).
From the rotation curve we derive a dynamical mass for the central 
region (3.3'' FWHM) of (3.9$\pm$1.6)$\times$10$^{10}$\solm, assuming  
a rotation
velocity of (290$\pm$60) km/s for the nuclear component.
The molecular mass
is then about $\sim$20\% of the dynamical mass in the nucleus.
This is at the upper end of what is expected for 
normal spiral galaxies (typical values are 5 -10\%; e.g. McGaugh \& de
Blok, 1997).

{\it The $\frac{N_{H_2}}{I_{CO}}$-conversion factor:}
Given estimates of the dynamical and stellar masses in the nucleus we are able to test the
$\frac{N_{H_2}}{I_{CO}}$-conversion factor. This factor was calculated for
molecular clouds in our Galaxy (e.g. Strong et al. 1987, van Dishoeck \& Black 1987 and
references therein) and
seems to be constant within a factor of two 
for ensembles of molecular clouds in other galaxies 
(e.g. Xie, Young, Schloerb 1994, Eckart et al. 1990, 1991, Eckart 1996,
Wilson \& Reid 1991, Tacconi et al. 1997, Solomon et al. 1997, but see also 
Maloney and Black 1988).
The standard conversion factor (2.0$\times$10$^{20}$$\frac{cm^{-2}}{K kms^{-1}}$;
Strong et al. 1987)
results in a molecular gas mass for the nuclear component of 
M$_{H_2}$=(7.5$\pm$1.5)$\times$10$^9$\solm.
The total dust mass is only a few times 10$^7$\solm ~(Eckart et al. 1994)
and can be neglected in the following.
We used the dynamical mass
of M$_{DYN}$=(3.9$\pm$1.5)$\times$10$^{10}$\solm ~derived from the rotation 
curve and the stellar mass of the 
starburst of M$_{ST}$=(1.9$\pm$0.7)$\times$10$^{10}$\solm (see section 4).  
Correcting for helium (36 \% of the molecular gas mass) we can now test the
conversion factor via
\\
\begin{equation}
\frac{M_{DYN} - M_{ST} - 0.36 M_{H_2}}{M_{H_2}} = 
\frac{M_{DYN} - M_{ST}}{M_{H_2}} - 0.36 = 2.3 \pm 1.4 
\end{equation}
\noindent
We find the \nhico-conversion factor to agree within the errors with the
canonical value of 2$\times$10$^{20}$$\frac{cm^{-2}}{K km s^{-1}}$ (Strong et
al. 1987). 
Remaining differences can easily be explained by uncertainties 
in our assumptions of the temperatures of the gas, the geometry assumed 
in calculating the dynamical mass, a higher fraction of warm molecular gas
and also by an old bulge population of late type stars which contributes 
approximately 2\% of the reddened stellar contribution to the K band flux
equals 1\% of the overall nuclear K band flux.
Such an old stellar population would have a mass of about 10$^{10}$\solm
(Thronson \& Greenhouse 1988)
and a luminosity of about 4$\times$10$^8$\solar.
The contribution of this old population cannot be substantially larger than
2\% of the reddened stellar K band flux
(or even account for all of it) without exceeding the total dynamical mass
(see section 5.2).
\\
Our result on I~Zw~1 and the findings for other lower redshift galaxies 
indicate that to within a factor of a few the average 
\nhico-conversion factor is valid in these extragalactic objects as well.
However, deviations may occur in low metallicity systems or in the presence
of substantial amounts of optically thin molecular gas.
In the metal-poor dwarf irregular galaxy IC~10 Wilson and Reid (1991)
find a conversion factor about twice as large as the Galactic value
in disagreement with predictions that the conversion factor should be a
strong function of metallicity.
In the SMC the conversion factor from CO line intensity to hydrogen
column density is found to be larger than the Galactic value and to
scale with the linear size of the structures as
$\chi_{SMC}$$\approx$9$\times$10$^{20}$
(R/10~pc)$^{0.7}$cm$^{-2}$(K~km~s$^{-1}$)$^{-1}$
(Rubio, Lequeux, Boulanger 1993).
A recent analysis of multiple CO transitions from the Cloverleaf quasar
by Barvainis et al. (1997) indicates that the conversion factor is probably
an order of magnitude smaller than the standard value due to the high emissivity of 
warm CO which is only moderately optically thick.
\\
The late type spiral shape of the I~Zw~1 host galaxy indicates normal metallicity
(see also optical line ratios in Phillips 1978)
and the $^{12}$CO and $^{13}$CO line ratios and the decomposition of the molecular
line emission into a nuclear and disk component (as discussed in Eckart et al. 1994)
indicates that most of the emission is due to warm optically thick (for the nuclear
component) and cold or subthermally excited molecular gas (for the disk component).
This indicates that models with an almost arbitrarily low conversion factor
due to a high emissivity of warm (optically thin or at least only moderately optically 
thick) molecular gas probably do not apply in the case of I~Zw~1.

\section{NEAR-INFRARED EMISSION FROM THE NUCLEUS AND THE HOST GALAXY}

\subsection{Continuum Images}

The host galaxy disk is clearly detected in the optical, NIR and in 
its molecular line emission (see Fig.1, 2 and 7).
There are two spiral arms originating east and west from the nucleus.
The western spiral arm is detected in the V, R and H band 
(Bothun et al. 1984, Smith et al. 1986, McLeod \& Rieke 1995) and 
in the $^{12}$CO (1-0) line emission (this paper).
The knotty structures in this spiral arm have been interpreted as HII regions 
(Smith et al. 1986), so the emission from this arm is expected to come 
mainly from newly formed stars.
The eastern spiral arm is obvious in the V and I band HST images
of I~Zw~1 (HST archive, Proposal 2882, P.I. Westphal, and K.D. Borne, 
private communication).
From the CO map, we see that the material for
star formation in the galaxy disk is more concentrated in spiral arms.
The western spiral arm is located at larger separations
from the nucleus than the eastern arm.
Both spiral arms are also evident in our IRAC2 H band map (Fig.7), but less 
pronounced in the K band data. Both images however clearly show the extended
central disk as well as the western companion and the foreground star.

\subsection{The Nuclear Near-Infrared Spectrum}

We present an H and K band (1.58 - 2.40 $\mu$m) NIR-spectrum of the nuclear region
(3.0'' aperture) with a spectral resolution of about
R=1000 for the H band and R=750 for the K band (see Fig.8).  The nuclear
NIR spectrum of I~Zw~1 is dominated by emission from the AGN with only a small 
contribution from a stellar population.
The spectrum shows a flat continuum indicating high extinction and dust 
emission as one would expect for a QSO. The spectrum is dominated by strong 
hydrogen recombination lines and also clearly shows the coronal [SiVI] line. We also
detected the CO(6-3) overtone bandhead at 1.62 $\mu$m which is due to absorption in
stellar atmospheres. 
All these lines originate within the central 2'' or are even
spatially unresolved at our resolution of about 1''.

\subsubsection{Emission Lines in the Near-Infrared}

The hydrogen recombination lines are produced in the BLR and NLR of the AGN as well as in the
HII regions around hot stars. To separate these two possibilities one can look at
the spatial distribution of the emission (e.g. NGC~7469, Genzel et al. 1995). The most 
prominent line in the K band nuclear spectrum of I~Zw~1 is the Pa$\alpha$ line at a rest 
wavelength of $\lambda_o$=1.875$\mu$m. Its maximum flux density
is 2/3 of the continuum flux density. Br$\gamma$ and Br$\delta$ are
two other H-recombination lines detected in the nucleus (see Fig.8).
The FWHM of Pa$\alpha$ is 1000 km/s comparable to the FWHM of Br$\gamma$ at our S/N ratio.
The FWZP is about 6000 km/s
although the line shape could be affected by an incomplete atmospheric correction, as there
is a strong atmospheric absorption feature nearby. This leads to an uncertainty of about 15\% in
the FWZP.
The line ratio of Pa$\alpha$ to Br$\gamma$ estimated from both the 
peak fluxes and integral line strengths is 11:1, close to the 12:1
that one would expect for standard HII-regions with case B recombination 
(N$_e$=10$^4$cm$^{-3}$ and T=10$^4$K; Osterbrock 1989).
The hydrogen recombination
line ratios estimated from the R band spectra
of Hutchings \& Crampton (1990) also do not differ
significantly from the theoretical values. 
Based on these ratios we conclude that there is little or no 
extinction towards the very nucleus of the QSO. 
\\
\\
Other emission lines which are detected are the two coronal emission lines [SiVI] 
($\lambda$1.962$\mu$m) and [AlIX] ($\lambda$2.040$\mu$m) (see Table 3), which arise in
hot gas (T$_e$ $\sim$ 10$^5$-10$^6$K) ionized by the AGN. They are 
blueshifted by 1350 km/s relative to the hydrogen recombination lines. Such a blueshift has
also been observed in the [CIV] emission line at $\lambda$1549$\AA$ (Buson \& Ulrich 1990). 
This blueshift indicates that these lines could come from an AGN driven outflow.
Our observed line flux of [SiVI] is within the limit derived by Eckart et al. (1994).
The flux of Kawara et al. (1990) is higher than ours, possibly due to contamination
from the nearby H$_2$ S(3) line being included in their 0.01 $\mu$m wide filter.
Laor et al. 1997 find in the UV spectrum of I~Zw~1 that the blueshift for ion lines
increases with their ionization level. They interpret this as an outflowing component in the
BLR, where the ionization level increases with velocity. In addition they detected a weak UV
absorption system which indicates an outflow with a line of sight velocity of $\sim$ 1870 km/s.
Interestingly, Leighly et al. 1997 interpret oxygen absorption features near 1 keV detected in
ASCA X-ray spectra as a highly relativistic outflow from the innermost regions in other 
NLS1 (IRAS 13224-3809, 1H 0707-495, PG 1404+226). It might be interesting to 
investigate further, whether
the outflow in the IR, UV and the X-ray are physically related.

\noindent
{\it HeI/Br$\gamma$-ratio}
\\
The HeI(2.058$\mu$m)/Br$\gamma$-ratio is sensitive to the effective temperature of the 
ionizing source and to the electron density (Doyon et al. 1992, Shields 1993),
since both lines are supposed to arise around the ionizing source.
Since we do not detect the recombination line of HeI (2.058$\mu$m) in our K band nuclear
spectrum, 
we get an upper limit of $\sim$ 0.30 for the ratio using the 3$\sigma$ error as an upper 
limit for the HeI line flux.
In this case the effective temperature must be either below
38 000 K or above 50 000 K with an electron density of $\sim$ 10$^3$ cm$^{-3}$
(Doyon et al. 1992, Shields 1993, Lan\c{c}on \& Rocca-Volmerange 1996).
Since the QSO nucleus shows at short optical wavelengths 
a strong contribution of blue non-stellar power law emission (Barvainis 1990)
the effective temperature should be large and probably be of the order (or exceed)
50 000 K.

\subsubsection{Absorption Lines in the Near-Infrared}

The most prominent stellar absorption features are the CO bandheads in the H and K band.
These lines arise in the atmospheres of late type giants and supergiants which produce most
of the stellar NIR emission. Due to the redshift of I~Zw~1 we were not able to detect
the strong $^{12}$CO(2-0) bandhead at 2.29 $\mu$m, as this line was already shifted out of
our detector range. In the H band the $^{12}$CO (6-3) overtone bandhead is accessible,
however. This clearly detected feature indicates a prominent stellar contribution to the 
continuum light of the nucleus. We used the depth of this line to estimate the fraction 
of stellar flux in the H band due to late type giants and supergiants.
The observed line depth is about (3.5$\pm$0.5)\% of the continuum. This can be compared 
to $\sim$ 20 \% we typically expect for a population of GKM giants/supergiants which
dominate the H band light for stellar populations older than 10$^7$ yrs. Therefore, we
estimate that at least 18$\pm$3 \% of the total
flux in the H band is due to GKM giants/supergiants. Although these stars
dominate the stellar emission in the H band, about one third of the total stellar flux is 
contributed by stars of other spectral classes. Therefore we derive that 27$\pm$6 \% 
of the H band continuum in a circular 3'' aperture is of stellar origin.
\\
In the K band the stellar absorption lines of the CaI-triplet (at 2.26$\mu$m) are normally
present in a stellar population of GKM giants/supergiants but not as prominent as the
CO bandheads. These absorption lines are only detected close to their 
3$\sigma$-level in I~Zw~1. If we compare the line depth
of 2.4$\pm$0.7~\%  
to spectra of late type standard stars with a similar resolution (Ali et al. 1995) which  
have line depths of about 10 \%, we derive an upper limit of about 24 \% for the contribution
to the stellar emission from this kind of late type stars. 

\noindent
These estimates of the stellar contribution to the NIR light are consistent with other
values in the literature. Barvainis (1990) finds about 20 \% of stellar continuum flux in 
the K band and about 30 \% in the H band in his spectral decomposition. 
Via a NIR color decomposition Eckart et al. (1994) find a K band contribution of 20 \%
with an extinction of A$_V\approx$10$^{mag}$. Comparing our stellar H and K fluxes 
(magnitudes) to the approximate NIR H-K colors of 0.20 (Frogel et al. 1978) of an
unextincted stellar population we can confirm an extinction of about 10$^{mag}$.
The Sc colors can be taken as a reference to obtain first order information on the
reddening.  This is reasonable since the spread in the mean colors
for different galaxy types (e.g. Hunt et al. 1997) is small compared to our calibration errors.
The distribution of typical giant and dwarf colors - taken as extreme cases of major flux density
distributors in the infrared - indicate that the colors of most stellar systems are very
similar. As demonstrated via a spectral synthesis analysis 
in Schinnerer et al. (1997) this argument can also be extended to 
starburst populations.
\\
The QSO nucleus of I~Zw~1 is located within this extincted central stellar
component and the very low extinction in the line of sight towards the nucleus
is most likely due to the central outflow (see section 4.2.1).

\section{THE CIRCUMNUCLEAR STARBURST RING}

The presence of a molecular 
ring in the circumnuclear region of I~Zw~1 , the large stellar
contribution to the total nuclear light as well as the analysis given
by Eckart et al. (1994) indicate that there is a massive starburst 
in the circumnuclear region of I~Zw~1.
Although the central AGN 
could have a substantial effect on the ISM in the I~Zw~1 host galaxy
we concentrate in the present paper on an interpretation in which 
we demonstrate that most of the ISM and stellar disk properties 
can be accounted for by star formation activity.
We use the new data presented in this paper together with a 
starburst model in order to investigate the properties of the starburst.
The K band luminosity L$_K$, the bolometric luminosity L$_{bol}$,
the Lyman continuum luminosity L$_{Lyc}$ and the supernova rate $\nu_{SN}$ are
used as observational parameters that are fitted by the model to explain the 
starburst age, history and upper mass cut-off.
In this section we describe the starburst model, the derived the 
input parameters, and the results of the modeling.

\subsection{The Starburst Model}

To derive the properties of a starburst from the observed continuum and
line intensities we have used
the starburst code STARS (Kovo \& Sternberg 1997).
This model has been successfully applied to NGC 1808 (Krabbe et al. 1994, 
Tacconi-Garman et al. 1996),
NGC~7469 (Genzel et al. 1995), NGC~6764 (Eckart et al. 1996) and NGC~7552 
(Schinnerer et al. 1997). A description of the model can be found in the appendices of Krabbe 
et al. (1994) and Schinnerer et al. (1997) and also in Kovo \& Sternberg (1997).
The model is similar to other stellar population synthesis models
(Larson \& Tinsley 1978, Rieke et al. 1980, Gehrz, Sramek \& Weedman 1983,
MasHesse \& Kunth 1991, Rieke et al. 1993,
Doyon, Joseph \& Wright 1994) and includes the most recent stellar evolution
tracks (Schaerer et al. 1993, Meynet et al. 1994). 
\\
We assume power-law IMFs which vary as 
M$^{-\alpha}$ between a lower and upper mass cut-off, m$_l$ and m$_u$, with an index
$\alpha$=2.35 (Leitherer 1996, Salpeter et al. 1955). STARS has
as output observable parameters such as the bolometric luminosity L$_{bol}$,
the K band luminosity L$_K$,the Lyman continuum luminosity L$_{Lyc}$ and the
supernova rate $\nu_{SN}$, as well as the diagnostic ratios between these quantities:
L$_{bol}$/L$_{Lyc}$, L$_K$/L$_{Lyc}$ and 10$^{9}\nu_{SN}$/L$_{Lyc}$.
All three ratios are measures of the time evolution and the shape of the
IMF, with slightly different dependencies on $\alpha$ and m$_u$.
H-R diagrams representing the 
distribution of these luminosities are calculated. 
The total number of stars of different stellar type is also calculated. 
To derive a total stellar
mass we integrated over the IMF represented by this H-R diagram.
\\
\\
There is no straightforward way to measure the bolometric luminosity
of the central region of I~Zw~1. There are, however,
two methods to estimate L$_{bol}$, and both result in comparable
values that are in good agreement with the spectral decomposition by
Barvainis (1990).
{\it Firstly} we can use the the CO-FIR relation (Young and Scoville 1991),
  in order to obtain L$_{FIR}$ as an estimate of L$_{bol}$ .
From our  high angular resolution millimeter data we can derive a 
central molecular gas mass of m$_{H_2}$=7.5$\times$10$^9$\solm.
From Fig.7 in Young \& Scoville (1991) we
get a L$_{bol}$=6.3$\times$10$^{10}$\solar ,
and dust temperature in the range of 40 to 60~K.
{\it Secondly, } this can be compared to an estimate of 
L$_{bol}$ via the  well-known FIR-radio relation 
(Wunderlich \& Klein 1988).
This method is in good agreement with the result
using the IRAS fluxes as shown by  Wunderlich \& Klein (1988)
and summarized in Schinnerer et al. (1997). 
For I~Zw~1 we can then use the
5 GHz radio flux to calculate L$_{bol}$. 
In a comparison to NGC~1068, Eckart et al. (1994) have provided
evidence that most of the total 5 GHz radio flux can be attributed 
to the central starburst.
By assuming:
\\
\begin{equation}
L_{bol} [L_{\odot}]=1.34 \times 10^{-7}
\times (1.13 \times 10^{17} \times D[Mpc]^2 \times S_{5GHz}[mJy])^{0.791}
\end{equation}
\noindent
where S$_{5GHz}$ is the flux density at 5GHz leading to a 
L$_{bol}$=4.6$\times$10$^{10}$\solar (Schinnerer et al. 1997).
The two methods show a good agreement. 
\\
\\
L$_{Lyc}$ and L$_K$ were estimated via 
\\
\begin{equation}
L_{Lyc} [L_{\odot}]=3.57 \times
10^{19} \times F_{Br\gamma}[erg s^{-1} cm^{-2}] \times D[Mpc]^2
\end{equation}
\noindent
and
\begin{equation}
L_K [L_{\odot}]=1.14 \times 10^4 \times D[Mpc]^2 \times S_K [mJy]
\end{equation}
\noindent
F$_{Br\gamma}$ is the Br$\gamma$-line flux
and S$_K$ is the K band flux density (Genzel et al. 1995). 
These quantities have to be extinction corrected. 
The supernova rate $\nu_{SN}$ is calculated as 
\\
\begin{equation}
\nu_{SN} [yr^{-1}]=3.1 \times 10^{-6} \times S_{5GHz}[mJy] \times D[Mpc]^2
\end{equation}
\noindent
(Condon 1992). The adopted values for L$_{Lyc}$ and L$_K$ are discussed in
the following section 5.2. All the luminosities as well as the 
resulting diagnostic ratios are listed in Table 4.

\subsection{Properties of the Nuclear Starburst}

If all the stellar K-band light were due to a very old stellar population
the corresponding mass (derived as demonstrated in e.g. Thronson \& Greenhouse 1988)
- reddened or dereddened - would exceed the overall dynamical mass determined in 
section 3.3 by 1 to 2 orders of magnitude. Combined with the other evidence provided in the
present paper and in Eckart et al. (1994) we conclude that a starburst 
(or at least enhanced star formation activity) takes place in the central 
region of I~Zw~1 and therefore contributes a major
fraction of the stellar near-infrared emission.
\\
We assumed two cases for the model starburst: (a) the QSO nucleus is
exclusively  powered by a strong starburst and
(b) only a fraction of the nuclear luminosity is due to a starburst which 
is located in the circumnuclear CO ring.

\noindent
If the QSO nucleus is powered by a strong starburst alone, the 
entire K band and Lyman continuum luminosity is due to that starburst. 
As shown in section 4.2.1 there is little or no extinction in the direction
of the nucleus for the Lyman continuum, so the same should be true for the 
K-band flux.
In this scenario, no model reproduced the observed diagnostic emission 
ratios at any metallicity. 
In all of them fitting both the Lyman continuum and K-band 
luminosity results in bolometric luminosities two or three orders of
magnitude larger than what is measured.
Thus we rule out this possibility for the origin of the nuclear 
luminosity in I~Zw~1.

\noindent
If we assume, however, that only 20\% of the K band flux is of stellar 
origin as it is suggested by the NIR spectrum (see section 4.2.2),
these stars are located within the central 2'' and are probably 
associated with the 
circumnuclear ring we found in the $^{12}$CO line emission. 
For this stellar component we adopt an extinction of
A$_V \approx$ 10$^{m}$ as derived by Eckart et al. (1994) and
discussed in section 4.2.2.
Since the fraction of the Lyman continuum luminosity which is due to a 
starburst is not known, we tried to find the fraction of the Lyman 
continuum luminosity that would give a reasonable fit for the 
diagnostic ratios and would still agree with 
the observations of the Br$\gamma$ flux. 
\\
This fit was obtained by maximizing 
a probability function proportional to exp(--$\chi^2$),
here $\chi^2$ is the sum of the 3 individual $\chi^2$-values
calculated from the predicted and measured diagnostic ratios and their
errors. 
L$_K$, L$_{bol}$, and $\nu_{SN}$ were kept fixed and L$_{Lyc}$
was varied.
The best fit is obtained if $\sim$ 0.5 \% of the observed Lyman 
continuum luminosity and therefore also about
$\sim$ 0.5 \% of the Br$\gamma$ line flux, used to derive the
Lyman continuum luminosity, is due to the starburst. This fraction of the 
Lyman continuum luminosity is highly extincted with respect to the rest 
of the emission (see section 4.2.1).
The best fit to the data is a decaying starburst with a decay-time of 
5$\times$10$^6$ yrs. The model burst started about 4.5$\times$10$^7$ yrs ago with a 
Salpeter IMF, and an upper mass cut-off of $\geq$ 90 \solm.
The present day stellar mass from this starburst is about
1.9$\times$10$^{10}$\solm.

\section{COMPARISION WITH STARBURST RINGS IN NEARBY GALAXIES}

Starburst rings are observed in a variety of galaxies (Buta \& Combes 1996, 
Maoz et al. 1996).
These rings are thought to be formed due to the gravitational interaction of 
the stars and the gas. They are detected in
the mid-infrared continuum, near-infrared colors, molecular gas line emission as
well as in H$\alpha$ line emission. The overall structure of these rings is 
not entirely smooth, so they could also be formed by two spiral 
arms tightly twisted around the nucleus.
\\
To test if the properties of the starburst ring in I~Zw~1 are unique or more 
common we made a comparison (Table 5) to 
the starburst rings in the HII galaxy NGC~7552 (Schinnerer et al. 1997)
and the Seyfert~1 NGC~7469 (Genzel et al. 1995), since these two rings were
studied over the same wavelength range, and using the same starburst model. 
I~Zw~1 is six times further away than NGC~7552, and three times more distant than
NGC~7469. In the following comparison we refer to the data given in Table 5.
\\
\\
In NGC~7552 Schinnerer et al. (1997) were able to study the properties of the
starburst ring in detail, since it is located at a distance of only 20~Mpc.
The size of the ring is about 0.8~kpc, comparable to
the ring detected in I~Zw~1.
The starburst is not smoothly distributed over the ring, rather there 
are regions of enhanced and lower star formation.
The 10$\mu$m dust emission is also not smoothly distributed along
the ring, suggesting a variation in dust mass or temperature.
The ages of the different regions, 
as derived in Schinnerer et al. (1997), differ by about a factor of two.
These individual starburst regions in the ring are about 
1.5$\times$10$^7$yrs old (with a decay time of 5 Myrs) and appear to have a high 
upper mass cut-off. 
The overall bolometric luminosities of the two starburst rings are similar. 
The difference in the Lyman continuum and K band luminosities can be
explained by the different ages, since with a decay-time of 5~Myrs, the 
Lyman continuum decreases as more and more of the massive stars die.
The K-band flux, then increases since more K and M supergiants/giants form. 
Therefore, the overall properties of the circumnuclear starburst in I~Zw~1 
are similar to the burst in NGC~7552, except that it may 
be 2 to 3 times older in I~Zw~1.
\\
\\
NGC~7469 harbors a Seyfert~1 nucleus which is surrounded by two tightly wound starbursting
spiral arms (Tacconi et al. 1997), previously thought to be a starburst ring. 
The properties of the starburst are similar to the burst in NGC~7552. 
Since the Br$\gamma$ emission in NGC~7469 is knotty the starburst probably varies
in age or strength between the different star formation sites.
The ''ring'' is a bit larger than in NGC~7552, and is the size of that in I~ZW~1. 
The bolometric luminosity in NGC~7469 is three times larger than that 
in NGC~7552. The two circumnuclear starbursts in NGC~7469 and I~Zw~1 are comparable 
in size and also show similar starburst properties.
As in the case of NGC~7552, the differences in Lyman continuum versus 
K band luminosity can be explained with different ages of the two
starbursts, and the correspondingly different evolutionary stages of the 
stellar populations.
\\
\\
Summarizing, we find that the properties of the starburst ''rings'' are
similar in all three galaxies. The differences in the total bolometric 
luminosity might be linked to the internal structure of the rings, 
and therefore the fueling of the individual starburst regions within the ring. 
The starburst in the ring of I~ZW~1 is about three times older than those in the
nearby objects NGC~7469 and NGC~7552. 
From this comparison it would seem that the molecular ring we 
detected in the $^{12}$CO(1-0) line emission is hosting a starburst 
quite similar to those found in other circumnuclear rings.
This could indicate that a fraction of the high luminosity observed for 
QSOs and Seyfert's is due to a circumnuclear starburst in the centers
of their host galaxies, and that the AGNs are not alone responsible 
for the overall energy output in the optical/infrared.
The contribution of the star formation activity to the bolometric
luminosity can range from only about 10 \% as in the case of I~Zw~1 to
about 50 \% as observed for NGC~7469 (Genzel et al. 1995).

\section{STAR FORMATION ACTIVITY IN THE I~ZW~1 HOST GALAXY}

In order to investigate the star formation activity in the spiral arms
of the I~Zw~1 host galaxy we use the population synthesis model as
described in section 5.1, and compare the optical-infrared broad band 
spectra of the northwestern spiral arm in the I~Zw~1 host galaxy
and the western companion, to those of spiral galaxies, ellipticals and 
the disk and nucleus of NGC~7469 (Fig.9).
From both the model calculations, and the bluer optical-infrared spectral energy
distribution we find indications for enhanced star formation on the arm.
This result is in agreement with the finding of Hutchings \& Crampton (1990).
From the optical colors and the absence of a definite Mg~IIb band 
feature in the I~Zw~1  disk these authors inferred that the host galaxy 
generally has an early type stellar population.
For the companion we find V-K colors which are within their errors similar to those of 
ellipticals or spirals. 
We find that both the nucleus and the disk of I~Zw~1 show properties of
high redshift QSOs and their host galaxies.
The indication of enhanced starburst activity from our population synthesis 
and the presence of significant amounts of molecular gas in the 
investigated area on the northwestern spiral arm 12~kpc off the nucleus
suggests that star formation and not scattered light from the QSO nucleus is 
responsible for the blue disk colors.

\subsection{Population synthesis}
Although the observations of the I~Zw~1 galaxy disk are of moderate S/N, it is possible to 
derive useful values for luminosities and the diagnostic ratios needed for 
the population synthesis model calculations.
To get a consistent estimate of the luminosities we calculated
all of them in  5'' diameter circular apertures
centered 9'' W and 5'' N of the nucleus right on the northwestern arm.
We did not correct for reddening, since the average extinction in galaxy disks 
is often found to be low (A$_V$$\approx$1 ; Frogel et al. 1978).
\\
To estimate the supernova rate $\nu_{SN}$ we compared the 1.4~GHz 
flux of 6.22$\pm$0.35~mJy measured by Barvainis \& Antonucci (1989)
in a 9'' beam to the flux of 7.30$\pm$0.45~mJy in the NRAO VLA Sky 
survey (NVSS) at 1.4~GHz (Condon et al. 1996) with a beam size of 45''. 
We assume that the 1.1mJy difference in the two flux measurements is an 
upper limit for the emission from the disk with roughly the same 
size as the NVSS beam. 
To get the flux at 5~GHz we use a spectral index $\alpha$=-0.88 as 
derived by Barvainis \& Antonucci (1989) for I~Zw~1, and obtain a flux 
density value of 0.38mJy in a 45'' beam.
Since the NVSS beam encompasses most of the I~Zw~1 host galaxy this value would correspond to 
an overall supernova rate of 0.07 yr$^{-1}$ which is a few times the estimated 
overall supernova rate in the Milky Way of (0.025$\pm$0.006) yr$^{-1}$
(Tammann, L\"offler, Schr\"oder 1994).
The  flux density value of 0.38mJy in a 45'' beam corresponds to 
5$\times$10$^{-6}$Jy in a 5" beam at 5~GHz. 
This flux can be compared to the estimate of 0.5~mJy noise in a 45" beam 
at 1.4~GHz just off the disk (2' separation from the nucleus) in the 
NVSS map.
This value results in 2$\times$10$^{-6}$Jy in a 5" beam at 5~GHz.
Using 5$\times$10$^{-6}$Jy in a 5" beam as an upper limit,
we obtain a supernova rate $\nu_{SN}$ of 0.001 yr$^{-1}$.
\\
We derived the Lyman continuum luminosity by the 
H$\beta$ flux of approximately 10$^{-15}$erg~s$^{-1}$~cm$^{-2}$ 
over a $\sim$3''$\times$8'' slit aperture in the off-nuclear spectrum 
by Hutchings \& Crampton (1990), and a ratio of 36:1 for H$\beta$:Br$\gamma$. 
The resulting Br$\gamma$ line flux is $\sim$ 2.3$\times$10$^{-17}$erg~s$^{-1}$~cm$^{-2}$ 
in a 5'' aperture.
This is clearly below our upper limit of 
3.3$\times$10$^{-15}$erg~s$^{-1}$~cm$^{-2}$
(3$\sigma$) derived from our spectroscopic imaging data. 
The derived value leads to L$_{Lyc}$=4.9$\times$10$^7$\solar. 
\\
From the IRAC2 ~K band image we obtain an average value for the flux 
density of 1.2$\times$10$^{-4}$Jy in a 5''~diameter aperture. 
This translates into a K band magnitude of 16.8$^{mag}$, and 
gives a K band luminosity of L$_K$=8.1$\times$10$^7$\solar.
\\
To get an estimate of the bolometric luminosity we used the CO-IR 
relation as described in section 5.1. From our 5'' channel maps 
with a spectral resolution of 10 km/s we get an 
I$_{CO}$=3.35~K~km/s with $\triangle$v$\sim$40~km/s and S$\sim$20~mJy. 
Correcting for the beam filling factor assuming a 15'' diameter disk,
this converts into a molecular H$_2$ mass of about 
3.1$\times$10$^8$\solm, and leads to L$_{bol}$=5.0$\times$10$^9$\solar
(assuming a dust temperature in the range of 30 to 40~K).
\\
Assuming solar metallicity we can now use these values and their
diagnostic ratios to determine the properties of a single underlying
starburst (see Fig.8 in Genzel et al. 1995, or Fig.10 in Schinnerer et al. 1997).
The diagnostic ratios calculated from the above values indicate a young 
decaying starburst with an age of $\sim$1.3$\times$10$^7$yrs and 
a moderate upper mass cut-off of m$_u$ $\sim$ 15 \solm.
The ratios are also in agreement with an old ($\sim$10$^{10}$ yrs),
constant starburst and an upper mass cut-off of 20 to 30 \solm.
In any case our estimates of the diagnostic ratios indicate 
active star formation in the northwestern arm of 
the I~Zw~1 host galaxy disk.

\subsection{Broad band spectra}

From our IRAC2 ~H and K (Fig.7) band images as well as the V and I 
band HST images (see Fig.2 (Plate~1))
we obtained flux densities (and the corresponding
magnitudes) in  5'' circular apertures centered on the northwestern
arm (9'' W and 5''N), on the western companion,
and on the nucleus of I~Zw~1 itself.
We compared these data to the mean spectra of spiral galaxies,
ellipticals, and the disk and nucleus of NGC~7469 (Fig.9).
\\
The spiral galaxy data for representative RC3 types 0 to 2 (S0 - Sab)
and 6 to 8 (Scd - Sdm) were taken from de Jong (1996).
We took the data on ellipticals from  Goudfrooij et al. (1994)
for the B, V, and I band and from Silva \& Elston (1994)
for the J, H, and K band.
The galaxies NGC~3377, NGC~3379, and NGC~5813 are listed by both groups. 
For these three objects the H band magnitudes
were derived using the mean NIR colors of the other ellipticals in the 
same sample.  In Fig.9 we display the combined data.
The data on the disk of NGC~7469 are taken from 
Kotilainen and Ward (1994) measured in a circular annulus
of 6'' to 12'' diameter centered on the nucleus.
For the nucleus the data are from Kotilainen,  Ward, Williger (1993)
for V and I band, and Kotilainen et al. (1992) for H and K band
measured in a 3'' circular aperture centered on the nucleus.
\\
In addition we checked the relative calibration of the V and I band data
on I~Zw~1 using the visible (380--600~nm) and red (660--965~nm) 
spectra of the nucleus published by Hutchings \& Crampton (1990) 
and Osterbrock, Shaw \& Veilleux (1990).
Hutchings \& Crampton (1990) also provide an absolute calibration
of their optical spectrum and give a calibrated R band map. 
These spectra are almost continuous in wavelength range
and allow us to estimate the 
spectral shape of the nucleus ranging from the V to I band. 
The data are consistent with the  V and I band
calibration provided by the HST, indicating that the I band 
flux density dip in the 
broad band spectrum of the nucleus of I~Zw~1 is real.
This dip spectrally separates the blue optical QSO nucleus from
the red dusty circumnuclear environment in which the QSO nucleus is embedded.
\\
\\
Comparing the broad band spectra of spirals and ellipticals to those we obtained
on the off-nuclear regions in I~Zw~1, and the companion we find that in general,
the optical colors of the spiral arm and the companion are bluer than what
expected for normal (nearby) ellipticals and spiral galaxies.
Their spectra have shapes quite comparable to those of spirals and ellipticals.
The spectrum of the spiral arm (and
within the errors also the spectrum of the companion) is
as blue as or bluer than the spectrum of central 3'' of
the Seyfert~I galaxy NGC~7469, where a 
substantial fraction of the light is originating in a 
circumnuclear starburst (Genzel et al. 1995).
\\
An indication that host galaxies of QSOs at higher redshift may, 
in general, be bluer than what expected for normal (nearby) ellipticals and spiral galaxies
is given by several authors.
Hutchings and Neff (1997) find blue host colors for 2 out of 4
objects in the redshift range of z=0.6 to 3.0.
R\"onback et al. (1996) find blue hosts in 16 out of 21 
QSOs and quasars in  the redshift range of z=0.4 to 0.8.
The blue disk colors we find for the I~Zw~1 host
demonstrates that not only the spectral line emission of
the I~Zw~1 QSO nucleus shows properties of high-redshift QSOs 
(the blue shifted CIV line - Buson \& Ulrich 1990 - and the 
blue shifted SiVI line - this paper in section 4.2.1),
but also that the I~Zw~1 host galaxy has structural and 
broad band spectral properties similar to those of high 
redshift QSO host galaxies.

The bluer colors may indicate enhanced star formation, and are 
consistent with the interpretation of Hutchings \& Crampton (1990)
that the disk generally shows an early type population.
This conclusion is also consistent with the result of our 
starburst analysis in the previous section.
The star formation activity, as well as the presence of 
molecular material within the disk (section 3.1)
and on the arm, indicates that at least in this 
region -- 12~kpc off the nucleus -- it is most likely star formation
that is responsible for the blue disk colors.
An alternative explanation for the blue disk colors might be a
contribution of scattered light from the nucleus as in the case of 
3C~234 (Tran, Cohen and Goodrich 1995) or Cygnus~A (Ogle et al. 1997).
How large such a contribution might be has to be investigated by
high angular resolution spectro-polarimetry of the I~Zw~1 host 
galaxy.

\section{SUMMARY AND CONCLUSIONS}

Our study of the host galaxy of I~Zw~1 covers the wavelength 
range from the radio to the optical. 
We have mapped the distribution of molecular gas in the QSO host
in its $^{12}$CO(1-0) line emission. The total molecular gas 
mass is of the order of 9$\times$10$^9$\solm with more than 2/3 
of it located within the central 3.3''. 
This amounts to about 20\% of the dynamical mass in the center.
Modeling the disk dynamics suggests an inclination of 
38$^o$$\pm$5$^o$. The mass balance for the nucleus results in a 
\nhico-conversion factor in general agreement with 
2$\times$10$^{20}$$cm^{-2}K^{-1}km^{-1}s$ found
for molecular gas in our Galaxy and many nearby external galaxies.
\\
We have detected the spiral arms of the host in molecular 
line emission, and find a circumnuclear ring with a diameter 
of 1.8~kpc (1.5'').  NIR spectroscopy of the central 3'' has 
revealed the presence of a prominent stellar component in this region.
Combined with available radio data we conclude that a young, 
massive, and decaying starburst is associated with this circumnuclear ring.
The properties of this starburst ring are similar to the ones 
observed in other sources with nuclear activity. These
similarities indicate that these rings may be a common 
phenomenon and contribute a significant fraction to the 
central luminosity.
\\
\\
Our H and K band images as well as V and I band HST images, together with 
data taken from the literature, show that the star formation 
in the {\it disk} of I~Zw~1 is enhanced as well.
A comparison with broad band spectra of spiral galaxies, ellipticals 
and the nucleus and disk in NGC~7469 clearly suggests that the north 
western arm in the I~Zw~1 host galaxy is bluer than expected.
The enhanced star formation in the host may indicate tidal 
interaction with a companion that shows an energy distribution 
similar to ellipticals. The companion's optical spectrum is also 
marginally bluer than expected.
A similar behaviour is found for host galaxies of intermediate redshift QSOs
and quasars.
\\
\\
In summary we find that
both the nucleus and the disk of I~Zw~1 show properties of
higher redshift QSOs and their host galaxies.
Further studies of I~Zw~1 will therefore be useful to probe 
and understand properties of QSOs and their host galaxies at 
higher redshift.

\vspace*{1cm}

\acknowledgements

We are grateful to the ESO/MPG 2.2m telescope  
staff and staff of IRAM, especially D. Downes, for their support and hospitality.
We thank the staffs of the 3.5m telescope at Calar Alto and the 
WHT for their excellent support. We thank Th. Boller, R. Fosbury and N. Scoville
for their comments and the helpful discussions as well as our referee P. Maloney for
his comments and help to improve this manuscript. We are especially grateful to N. Thatte, L.E.
Tacconi-Garman and the rest of the MPE 3D team for taking the H and K band NIR
spectra. We also thank M. L\"owe and A. Quirrenbach for taking the H and K
band IRAC2 images at the ESO/MPG 2.2 telescope.

\clearpage

\begin{table}[htb]
\caption{$^{12}$CO(1-0) flux distribution in I~Zw~1}
\begin{center}
\begin{tabular}{lrrrrl} \tableline \tableline
 & FWHM & S$_{obs}$ & T$_{obs}$ & T$_{real}$ &Distribution \\ \tableline
nucleus & 3.3'' & 0.056 Jy & 0.0117 K & 0.551 K  & Gaussian\\ 
disk & 9.0'' & 0.022 Jy & 0.0046 K & 0.033 K & Gaussian\\
disk & 12.0'' & 0.022 Jy & 0.0046 K & 0.005 K & flat \\ \tableline \tableline
\end{tabular}
\end{center}
Observed flux density and main beam brightness temperature 
of the 30m flux measurements
as well as the source size corrected temperatures 
assuming sizes for the disk and nucleus as given.
\end{table}

\begin{table}[htb]
\caption{Molecular mass distribution in I~Zw~1}
\begin{center}
\begin{tabular}{lrrrrr}\tableline \tableline
 & $\triangle v$ & I$_{CO}$ & N(H$_2$) & A & M$_{H_2}$ \\ 
 & [km s$^{-1}$] & [K km s$^{-1}$] & [10$^{21}$cm$^{-2}$] & [kpc$^2$] & [10$^9$M$_{\odot}$] \\ \tableline
nucleus (3.3'') & 380  & 209.4  & 42 & 11.8 & 7.5\\
disk (9.0'') & 310 & 10.2 & 2.0 & 88.6 & 2.7\\
disk (12.0'') & 310 & 1.6 & 0.3 & 157.5 & 0.7 \\
\tableline \tableline
\end{tabular}
\end{center}
Derived mass of molecular hydrogen for the nuclear and disk. 
The velocity widths are measured
peak-to-peak for the disk component and FZW for the nucleus.
\end{table}

\begin{table}[htb]
\caption{Integrated NIR line fluxes of I~Zw~1}
\begin{center}
\begin{tabular}{lrrrr}\tableline \tableline
~~&wavelength &peak intensity &  flux  & [10$^{-18}$ W m$^{-2}$] \\  
  & [$\mu$m]    &[10$^{-15}$ W m$^{-2}$ $\mu$m$^{-1}$  & &  \\ \tableline 
~~& & 3D & 3D & FAST \\ \tableline 
Pa$\alpha$~&1.8751& 26.9$\pm$0.7   &   284$\pm$12 &  \\
Br$\gamma$~&2.1665&  2.4$\pm$0.7   &     26$\pm$8 &  \\
Br$\delta$~&1.9445&  1.5$\pm$0.7   &    3.7$\pm$1.8 &  \\
H$_2$ S(1)~&2.1218&  $<$0.15       &         $<$2.4 & 0.68$\pm$0.20\\
SiVI       &1.9615&  1.4$\pm$0.7   &   3.5$\pm$1.8 & $<$ 3.0 \\
AlIX       &2.040 &  0.7$\pm$0.7   &   3.2$\pm$3.2 &  \\
CO 6-3     &1.6185&  2.1$\pm$0.7   &  -4.6$\pm$1.5 & \\ 
\tableline \tableline
\end{tabular}
\end{center}
{\small The peak line intensities taken from the spectrum in Fig.8 and
integrated line fluxes in a 3'' aperture. The 3$\sigma$ errors
for both the peak and the integral values are given.
For comparison the line fluxes measured with FAST 
(Eckart et al. 1994) are listed in the 4th column as well. 
Within the uncertainties they are in agreement with the new 
data presented here.}
\end{table}

\begin{table}[htb]
\caption{Luminosities and diagnostic ratios for nuclear starburst models}
\begin{center}
\begin{tabular}{lccc}\tableline \tableline
 & model (a) & model (b) & fit \\ \tableline
S$_K$ [mJy] & 69.2 & 13.84 & \\
F$_{Br\gamma}$ [10$^{-18}$ Wm$^{-2}$] & 26 & $\sim$ 0.04 & \\
S$_{5GHz}$ [mJy] & 2.2 & 2.2  & \\
L$_{Bol}$ [10$^{10}$L$_{\odot}$] & 6.3   & 6.3   & 100 \\
L$_K$ [10$^{10}$L$_{\odot}$]     & 4.70  & 1.61  & 1.61 \\
L$_{LYC}$ [10$^{9}$L$_{\odot}$]  & 54.3  & 0.23  & 1.19 \\
$\nu_{SN}$ [yr$^{-1}$]            & 0.41  & 0.41  & 2.66 \\
& & & \\
$\frac{L_{bol}}{L_{Lyc}}$        & 1.2   & 273.9 & 840.4 \\
log($\frac{L_K}{L_{Lyc}}$)       & -0.06  & 1.84  & 1.13 \\
$\frac{10^9 \nu_{SN}}{L_{Lyc}}$  & 0.008 & 1.735 &2.236 \\
& & & \\
t [10$^7$ yr] & & 3.0 - 6.0 & 4.4 \\
m$_u$ [M$_{\odot}$] & & $\ge$ 90 & 120 \\
\tableline \tableline
\end{tabular}
\end{center}
Flux densities and line fluxes used to derive the luminosities and 
supernova rate of the circumnuclear starburst in I~Zw~1. 
The calculated diagnostic ratios and the ratios of a
good fit for model (b) are also given.
We assumed a distance of 244 Mpc and an A$_V$=10$^{mag}$ 
for model (b) and no extinction for model (a). 
The models as well as the predicted values are described in the text.
\\
\end{table}

\begin{table}[htb]
\caption{Comparision to other starburst rings}
\begin{center}
\begin{tabular}{lccccc}\tableline \tableline 
& NGC~7552 & NGC~7469 & I~Zw~1\\ \tableline
Type & LINER/H II & Seyfert 1 &QSO/Sey 1\\
Distance & 20~Mpc & 66~Mpc& 244~Mpc \\
Size & 17 kpc & 20 kpc & $\sim$ 30 kpc \\
Ring Diameter & 1 kpc & 1 kpc & 1.8 kpc \\
& & & \\
L$_{bol}$ [10$^{10}$\solar] & 3 & 10 & 6.3 \\
L$_K$ [10$^{9}$\solar] & 0.7 & 3 & 16 \\
L$_{Lyc}$ [10$^{9}$\solar] & 2.8 & 9.5 & 0.3 \\
$\nu_{SN}$ [yr$^{-1}$] & 0.1 & 0.4 & 0.4 \\
$\frac{L_{bol}}{L_{Lyc}}$ & 10 & 28 - 38 & 210 \\
$\frac{L_K}{L_{Lyc}}$ & 0.24 & $\sim$0.3 & 53 \\
$\frac{10^9 \nu_{SN}}{L_{Lyc}}$ & 0.034 & 0.049 & 1.33 \\
& & & \\
m$_u$ [\solm] & $\approx$ 100 & $\leq$ 100 & $\geq$ 90 \\ 
t$_{burst}$ [Myr] & $\sim$ 15  & $\sim$ 15  & $\sim$ 45  \\
type & dec. & dec. (?) & dec.\\ 
\tableline
\tableline
\end{tabular}

\end{center}
{\small Properties of the starburst rings in NGC~7552 taken from 
Schinnerer et al. (1997) and in NGC~7469 from Genzel et al. (1995). 
The data for I~Zw~1 are presented in this paper.}
\end{table}

\newpage

\begin{figure}
\begin{center}
\psfig{file=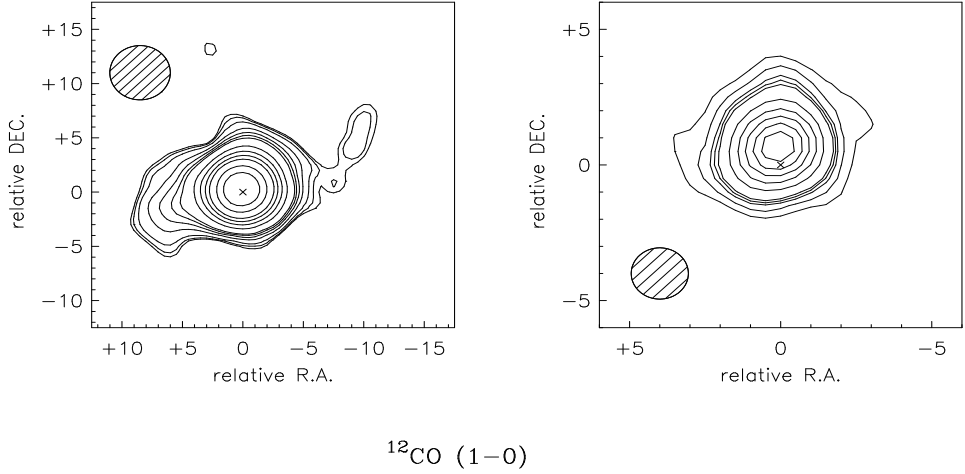,height=10.0cm,width=15.0cm,angle=-90.0}
\end{center}
\caption{
Integrated maps of the $^{12}$CO(1-0) line with an angular resolution of
5'' corresponding to 5.9~kpc (left, Fig.1a) and 1.9'' corresponding to 2.2~kpc 
(right, Fig.1b).
The contour levels are 0.077, 0.078, 0.080, 0.085, 0.090, 0.095, 0.10, 
0.125, 0.15, 0.175, 0.20, 0.25, 0.30, 0.35 Jy/beam for the
left map and 0.04, 0.06, 0.08, 0.09, 0.10, 0.15, 0.20, 0.25, 0.30, 0.35
Jy/beam for the right map.}
\end{figure}

\begin{figure}
\begin{center}
\psfig{file=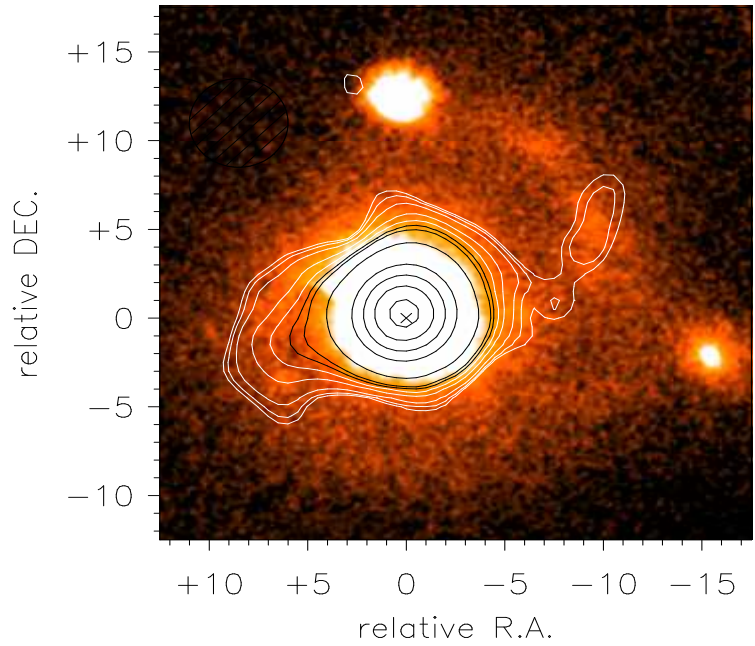,height=10.0cm,width=12.0cm,angle=-90.0}
\end{center}
\caption{
Integrated maps of the $^{12}$CO(1-0) line with an angular 
resolution of 5'' corresponding to 5.9~kpc. 
The contour levels are 0.077, 0.078, 0.080, 0.085, 0.090, 0.095,
0.100, 0.125, 0.150, 0.175, 0.200, 0.250, 0.300, 0.350, 0.400 Jy/beam.
Underlying the I band HST image of I~Zw~1 convolved to a resolution of
2'' for better comparison.}
\end{figure}

\begin{figure}
\begin{center}
\psfig{file=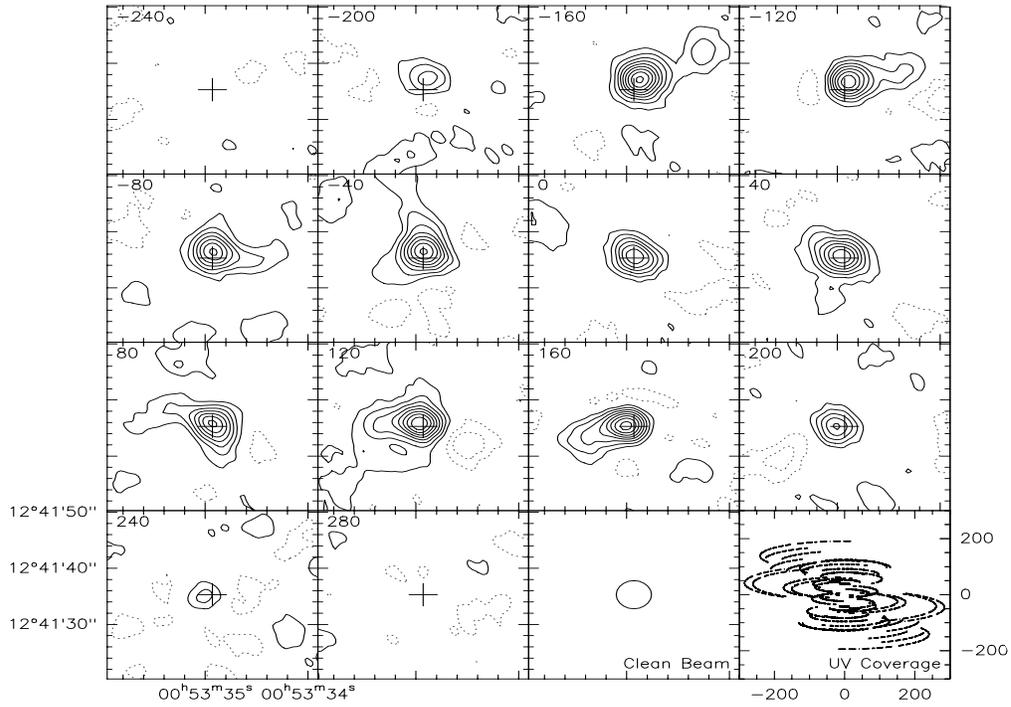,height=10.0cm,width=18.0cm,angle=-90.0}
\end{center}
\caption{
Channel maps of the $^{12}$CO (1-0) line emission with a spectral
resolution of 40 km/s and an angular resolution of 5''.
The contour levels are in steps of 5 mJy/beam.
Clean beam and uv coverage are plotted as well.
The central velocity of each channel with respect to the
systemic velocity is indicated in each image.
The extended emission of a rotating disk is clearly detected. }
\end{figure}

\begin{figure}
\begin{center}
\psfig{file=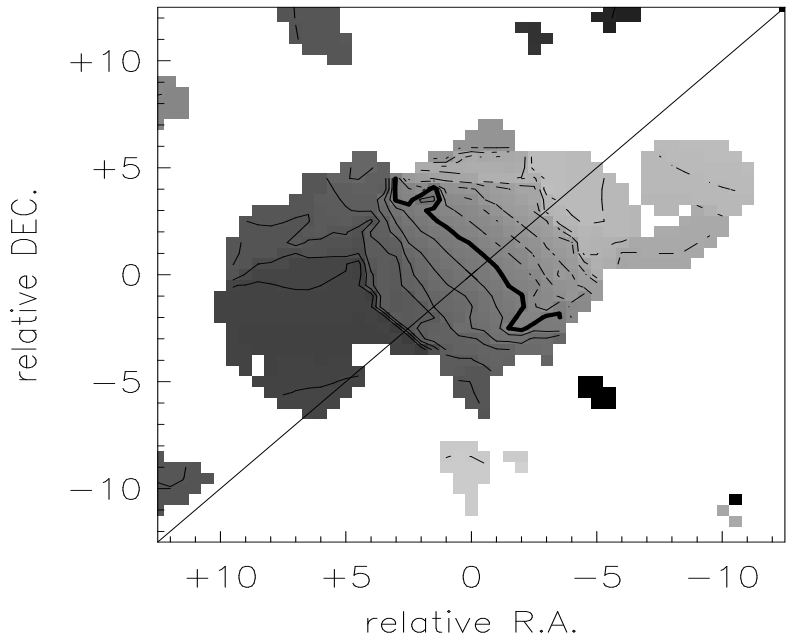,height=10.0cm,width=12.0cm,angle=0.0}
\end{center}
\caption{
Velocity field of the $^{12}$CO (1-0) line emission with an angular
resolution of 5''.
The contours are -240 km/s, -220 km/s, -200 km/s, ... , 0 km/s, ... , 240 km/s.
Light grey shading and dashed contour levels correspond to negative
velocities. The thick dark line is the 0 km/s contour.
For reference the kinematic major axis has been included as well.
}
\end{figure}

\begin{figure}
\begin{center}
\psfig{file=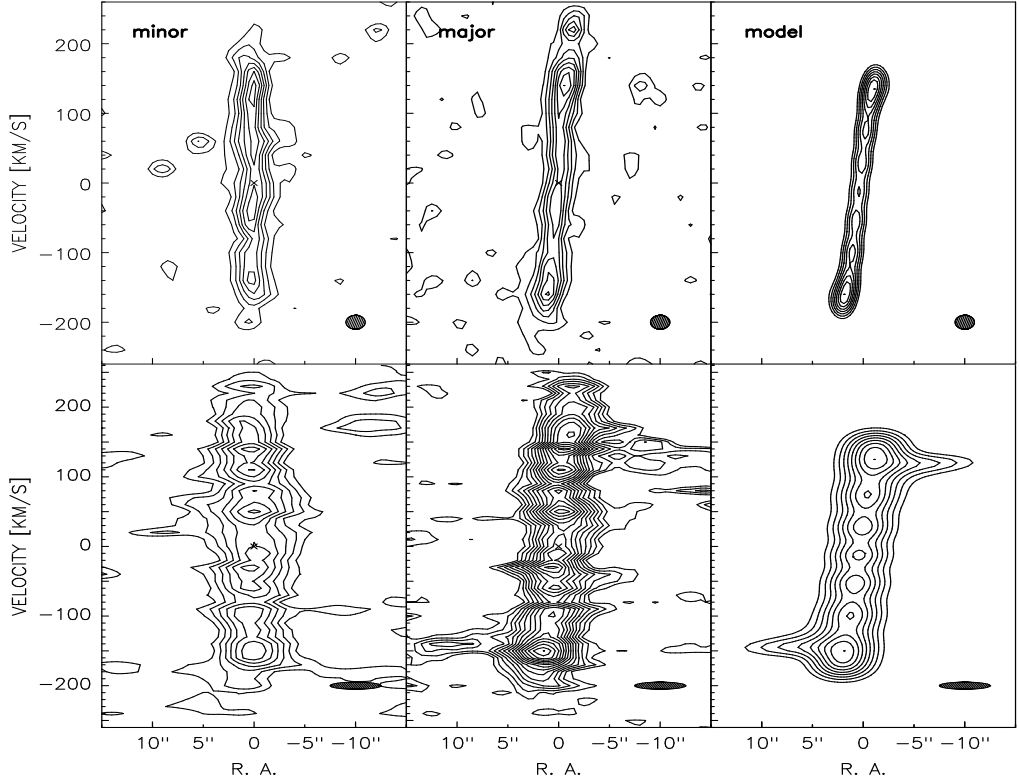,height=11.0cm,width=18.0cm,angle=-90.0}
\end{center}
\caption{
pv-diagrams of the $^{12}$CO(1-0) line emission along the
kinematic minor and major axis for the high angular 
resolution map (1.9'' FWHM; 20 km/s) (top)
and low angular resolution map (5'' FWHM; 10 km/s) (bottom).
In addition the modelled pv-diagrams for the major axis 
are plotted on the right.
The contour levels are equally spaced with separations
10\% starting at the 35\% (top and bottom left),
10\% starting at the 30\% level (top middle and right), 
5\% starting at the 30\% (bottom middle),
and 10\% starting at the 20\% (bottom right)
of the peak brightness.}
\end{figure}

\begin{figure}
\begin{center}
\psfig{file=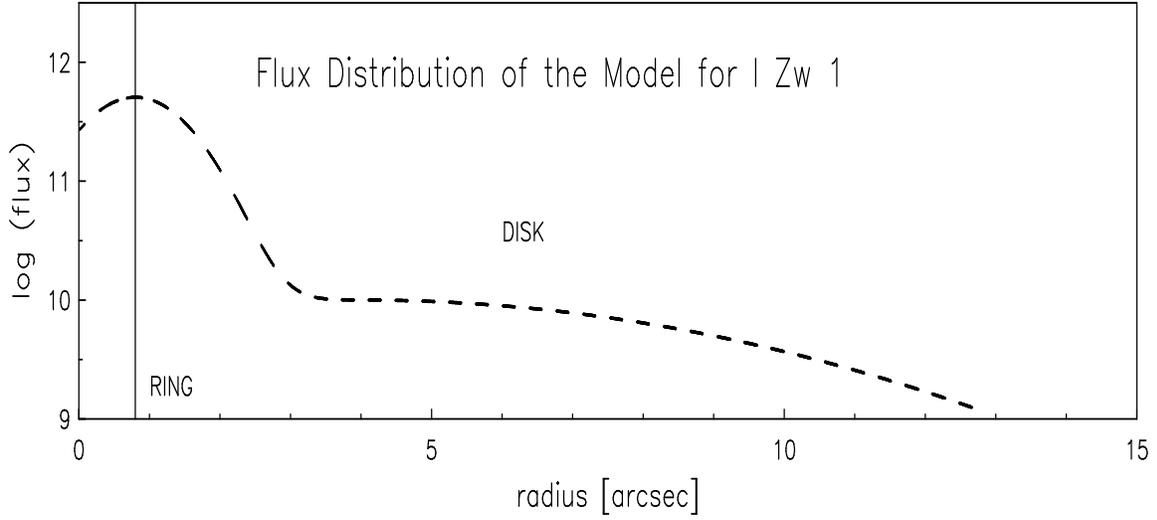,height=8.0cm,width=16.0cm,angle=-90.0}
\end{center}
\caption{
The $^{12}$CO(1-0) line flux distribution as a function of radius 
for the derived model.
A nuclear ring with a radius of 0.8'' is superposed on an underlying 
and much less luminous disk.}
\end{figure}

\begin{figure}
\begin{center}
\psfig{file=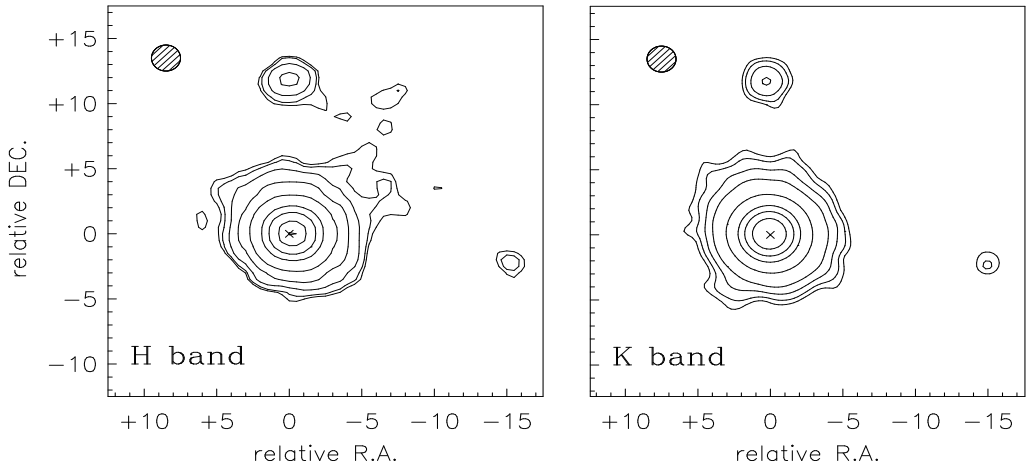,height=12.0cm,width=18.0cm,angle=-90.0}
\end{center}
\caption{
Maps of the H band (left) and K band (right) continuum emission.
Contour levels are 
1.2, 1.33, 1.7, 2.7, 5.3, 16.7, 33, 67, 93 \% for the H band
and 0.35, 0.47, 0.7, 1.1, 2,3, 3.5, 11.7, 23.5, 47, 95 \% 
of the peak brightness for the K band map.
As indicated by the beam in the upper left hand corner
the angular resolution in both images is 2''.
In both images the star to the north, the companion galaxy to the west
as well as the extended emission of the central disk are apparent.}
\end{figure}

\begin{figure}
\begin{center}
\psfig{file=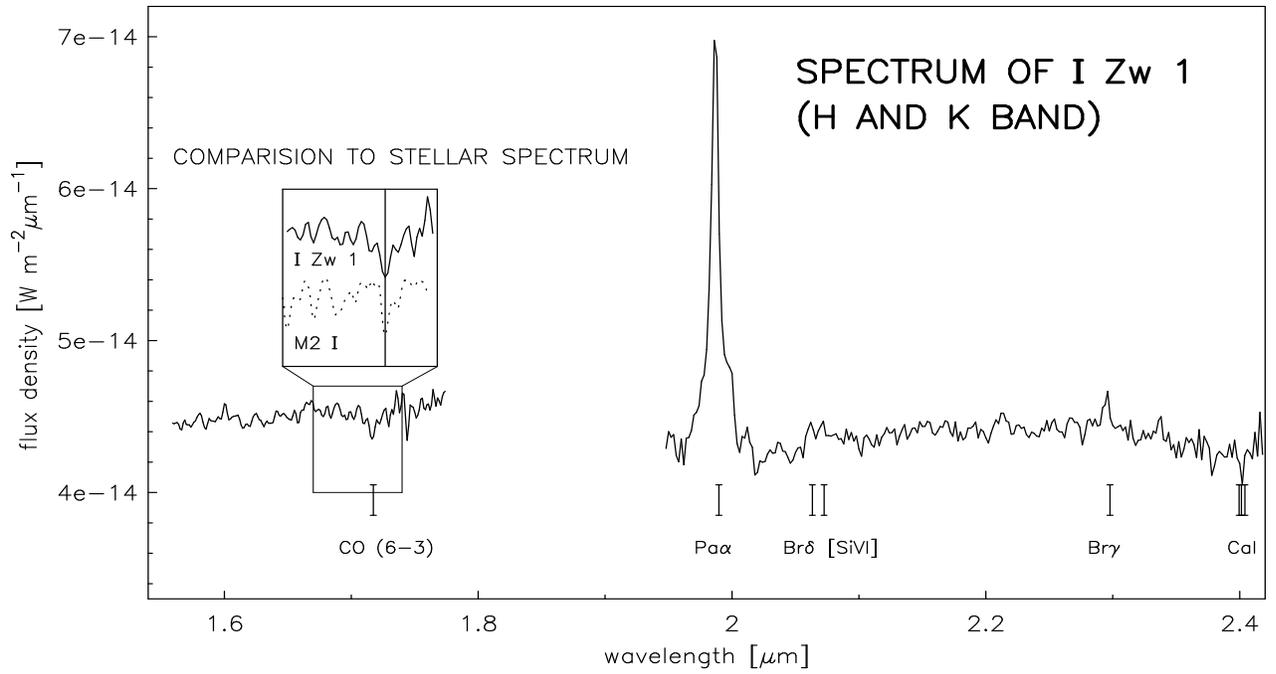,height=12.0cm,width=18.0cm,angle=-90.0}
\end{center}
\caption{
NIR-spectrum of I~Zw~1 in the H and K band (1.58 - 1.78 $\mu$m
and 1.95 - 2.40 $\mu$m).
The spectrum includes light from the inner 3'' of the nucleus.
For the CO(6-3) overtone a comparison to a stellar spectrum of 
an M2 I star convolved to the resolution of the measured I~Zw~1
spectrum (Dallier et al. 1996) is shown.
The positions of other prominent detected lines are indicated.}
\end{figure}

\begin{figure}
\begin{center}
\psfig{file=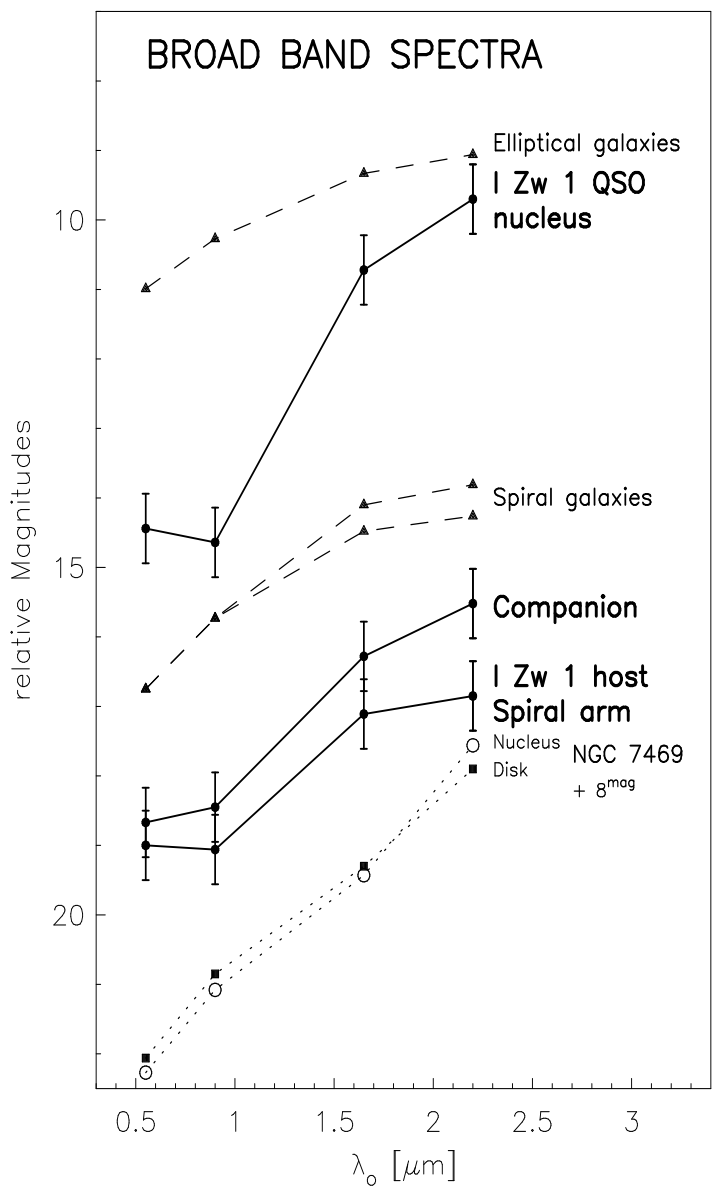,height=18.0cm,width=12.0cm,angle=0.0}
\end{center}
\caption{
Broad band spectra of the northwestern arm (9'' W and 5''N), 
on the western companion, and on the nucleus of I~Zw~1 itself.
For comparison we show mean spectra of spiral galaxies
(top curve: S0 - Sab; bottom curve: Scd - Sdm; de Jong 1996), 
ellipticals (Goudfrooij et al. 1994; Silva \& Elston
1994), and the disk and nucleus of NGC~7469 (Kotilainen and Ward 1994
and references therein). The spectrum of NGC~7469 was shifted by
+8$^{mag}$ for display purposes. The estimated errors on the I~Zw~1 
measurements (including positioning of aperture) are $\pm$0.5$^{mag}$.
See text for further information. 
}
\end{figure}

\end{document}